\documentclass{article}
\usepackage{spconf}
\usepackage{cite}
\usepackage{amsmath,amssymb,amsfonts,amsthm}
\usepackage{algorithmic}
\usepackage{graphicx}
\usepackage{textcomp}
\usepackage{xcolor}
\usepackage[table]{xcolor} %
\usepackage{flushend}
\usepackage{orcidlink}
\usepackage{multirow}
\usepackage{lipsum}
\theoremstyle{plain}
\newtheorem{thm}{Theorem}

\newtheorem{asm}[thm]{Assumption}

\newtheorem{example}[thm]{Example}

\usepackage{makecell}
\usepackage{svg}

\def\BibTeX{{\rm B\kern-.05em{\sc i\kern-.025em b}\kern-.08em
    T\kern-.1667em\lower.7ex\hbox{E}\kern-.125emX}}

\begin{document}
\ninept

\title{
{LipsAM: Lipschitz-continuous Amplitude Modifier for Audio Signal Processing and Its Application to Plug-and-Play Dereverberation}
}

\name{Kazuki Matsumoto, \ \ Ren Uchida, \ \ Kohei Yatabe\thanks{This work was partly supported by JST FOREST Program (Grant Number JPMJFR2330, Japan).}}
\address{Tokyo University of Agriculture and Technology, Tokyo, Japan \vspace{-3pt}}

\maketitle

\begin{abstract}

\vspace{-2pt}
The robustness of deep neural networks (DNNs) can be certified through their Lipschitz continuity, which has made the construction of Lipschitz-continuous DNNs an active research field.
However, DNNs for audio processing have not been a major focus due to their poor compatibility with existing results.
In this paper, we consider the amplitude modifier (AM), a popular architecture for handling audio signals, and propose its Lipschitz-continuous variants, which we refer to as \textit{LipsAM}.
We prove a sufficient condition for an AM to be Lipschitz continuous and propose two architectures as examples of LipsAM.
The proposed architectures were applied to a Plug-and-Play algorithm for speech dereverberation, and their improved stability is demonstrated through numerical experiments.
\vspace{-2pt}
\end{abstract}


\vspace{-2pt}
\section{Introduction}

Theoretical guarantees of the robustness and reliability of deep neural networks (DNNs) have been an active research topic.
Certifying robustness against adversarial examples \cite{tsuzuku_lipschitz-margin_2018,cisse_parseval_2017} and the stability analysis from a dynamic system perspective \cite{chen_neural_2018,chen_residual_2019} are some examples in this line of research.
One powerful tool for ensuring robustness is the Lipschitz continuity of DNNs.
Several approaches have been proposed to design Lipschitz-bounded DNNs, e.g., projection \cite{miyato_spectral_2018,ryu_plug-and-play_2019,li_efficient_2019}, regularization \cite{wangOrthogonalConvolutionalNeural2020,pesquet_learning_2021}, and structurally 1-Lipschitz layers \cite{li_preventing_2019,casado_cheap_2019,trockman_orthogonalizing_2020,prach_1-lipschitz_2024,meunier_dynamical_2022,ducotterd_improving_nodate}.
Such techniques have been applied to design DNN architectures, including recurrent neural networks \cite{erichson_lipschitz_2021}, self-attention \cite{kim_lipschitz_2021}, and transformers \cite{qi_lipsformer_2023}.
Computing the Lipschitz constant of a trained DNN is another important subject \cite{virmaux_lipschitz_2018,khromov_fundamental_2024}.

Although interest in Lipschitz-continuous DNNs has grown across many fields, audio signal processing has benefited less from these advances.
This is mainly because typical architectures for audio signals are often incompatible with existing theoretical frameworks.
For example, DNN-based time-frequency-domain processing, which transforms audio signals using the short-time Fourier transform (STFT) and then processes the obtained spectrograms, often modifies only the amplitude of the spectrograms \cite{narayanan_ideal_2013,hershey_deep_2016}.
In this case, the entire system is generally not Lipschitz continuous, even when the DNN that modifies the amplitude is Lipschitz continuous.
To ensure Lipschitz continuity in such audio-specific architectures, 
additional constraints or tailored designs are required.

In this paper, we focus on a DNN that modifies the amplitude of the spectrogram, which we call the amplitude modifier (AM), and we prove a sufficient condition for an AM to be Lipschitz continuous.
We also propose two Lipschitz-continuous AMs according to the obtained condition and name them \textit{LipsAM}.
Our contribution can be summarized as follows:
(1) We prove the condition for Lipschitz-continuous AM;
(2) We propose two Lipschitz-continuous AMs;
(3) We provide the bounds on their Lipschitz constants; and 
(4) We apply the proposed DNN architectures to a Plug-and-Play (PnP) method \cite{chan_plug-and-play_2017,ryu_plug-and-play_2019,pesquet_learning_2021} for speech dereverberation.

\textbf{Notation:} $\mathbb{R}_+$, $\mathbb{R}$, and $\mathbb{C}$ denote the sets of nonnegative, real, and complex numbers, respectively, and $\mathbb{F}\in \{\mathbb{R},\mathbb{C}\}$.
Lower bold letters represent vectors, %
e.g., $\mathbf{z} = (z_n)_{n=1}^N \in \mathbb{C}^N$, and
capital bold letters denote matrices, e.g., $\mathbf{A} = (a_{ij})_{i=1,\, j=1}^{I,\,J} \in \mathbb{C}^{I\times J}$.

\section{Preliminary}

\subsection{Lipschitz Continuity of DNNs}

A mapping $f:\mathbb{F}^N\to\mathbb{F}^{N}$ is said to be \textit{$L$-Lipschitz continuous} if there exists a constant $L \geq 0$ such that
\begin{align}
\forall \mathbf{x},\  \mathbf{y}\in\mathbb{F}^N,\quad\|f(\mathbf{x}) - f(\mathbf{y})\|_2 \leq L \|\mathbf{x} - \mathbf{y}\|_2,
\end{align}
and the smallest $L$ is called the \textit{Lipschitz constant}, $\mathrm{Lip}(f)$, given by
\begin{equation}
\mathrm{Lip}(f) = \sup_{\mathbf{x}\neq\mathbf{y}\in\mathbb{F}^N} \left(\frac{\|f(\mathbf{x}) - f(\mathbf{y})\|_2}{\|\mathbf{x} - \mathbf{y}\|_2}\right),
\end{equation}
where $\|\cdot\|_2$ is the $\ell_2$-norm.
Lipschitz continuity and smallness of the Lipschitz constant indicate robustness to the input perturbations.
Many theoretical guarantees of DNNs rely on the assumption of their Lipschitz continuity \cite{
tsuzuku_lipschitz-margin_2018,cisse_parseval_2017,chen_neural_2018,chen_residual_2019,ryu_plug-and-play_2019,pesquet_learning_2021}, and
several techniques have been proposed to construct Lipschitz-continuous DNNs \cite{miyato_spectral_2018,ryu_plug-and-play_2019,li_efficient_2019,wangOrthogonalConvolutionalNeural2020,pesquet_learning_2021,li_preventing_2019,casado_cheap_2019,trockman_orthogonalizing_2020,prach_1-lipschitz_2024,meunier_dynamical_2022,ducotterd_improving_nodate,erichson_lipschitz_2021,kim_lipschitz_2021,qi_lipsformer_2023}. %

\begin{figure}[t]
    \centering
    \includegraphics[width=0.99\linewidth]{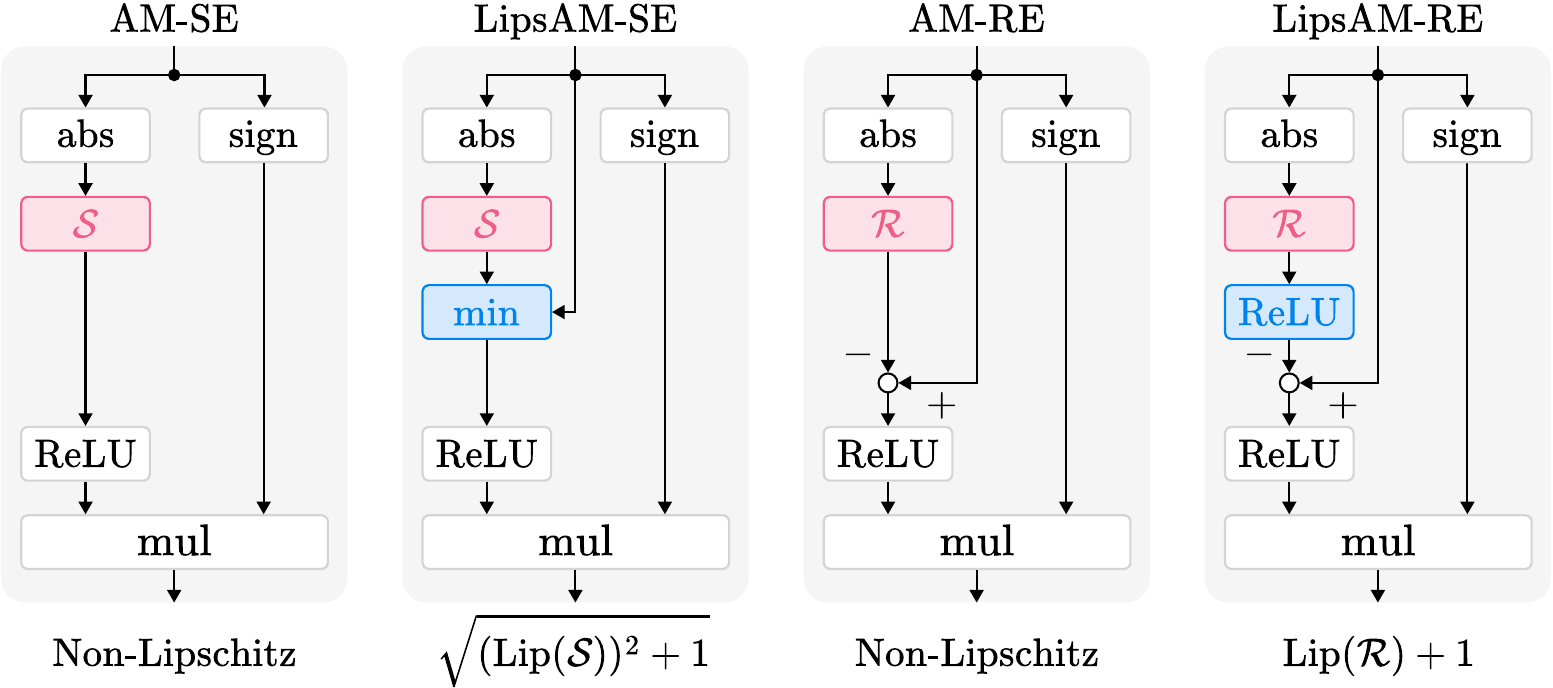}
    \vspace{-7pt}
    \caption{Amplitude-Modifying DNNs. 
    Red layers are trainable, blue layers are introduced to enforce Lipschitz continuity. 
    Layers ``mul'' represent the element-wise multiplication.
    While the popular architectures, AM-SE and AM-RE, are generally not Lipschitz continuous,
    Lipschitz constants of our LipsAM-SE and LipsAM-RE can be bounded by $\sqrt{(\mathrm{Lip}(\mathcal{S}))^2+1}$ and $\mathrm{Lip}(\mathcal{R})+1$, respectively.}
    \label{fig:architecture}
    \vspace{-5.2pt}
\end{figure}

\subsection{Amplitude Modifier (AM)}

In audio applications, signals are often processed in the STFT domain.
Let a signal in the STFT domain be vectorized and denoted as $\mathbf{z}\in\mathbb{C}^N$.
Typical DNN architectures process this signal by modifying its amplitude, $|\mathbf{z}|\in\mathbb{R}_+^N$, while the phase-related components, $\mathrm{sign}(\mathbf{z})\in\mathbb{C}^N$, remain intact,
where the absolute value and the complex-valued sign function ($\mathrm{sign}(z) := z / |z|=\exp(\mathrm{i}\,\arg z)$ if $z \neq 0$ and $0$ otherwise) are applied element-wise \cite{narayanan_ideal_2013,hershey_deep_2016}.
Here, we call such a DNN \textit{Amplitude Modifier (AM)}:
\begin{equation}
    \text{AM :}\quad\mathcal{D}_\mathcal{A}(\mathbf{z}) = \mathcal{A}(|\mathbf{z}|)\odot \mathrm{sign}(\mathbf{z}),
    \label{eq:amplitude}
\end{equation}
where $\mathcal{A}:\mathbb{R}_+^N\to\mathbb{R}_+^N$ is the DNN trained for processing, and $\odot$ denotes the element-wise multiplication.

In this paper, we consider two types of AM.
The signal estimator (AM-SE), $\mathcal{D}_\mathcal{S}$, directly estimates the amplitude spectrogram of the clean signal using $\mathcal{S}:\mathbb{R}_+^N \to \mathbb{R}^N$, and the residual estimator (AM-RE), $\mathcal{D}_\mathcal{R}$, estimates the residual component using $\mathcal{R}:\mathbb{R}_+^N \to \mathbb{R}^N$ and then subtracts it from the input, as follows:
\begin{align}
\text{AM-SE :}\quad&\mathcal{D}_{\mathcal{S}}(\mathbf{z}) = (\mathcal{S}(|\mathbf{z}|))_+ \odot \mathrm{sign}(\mathbf{z}),
\label{eq:signal}\\
\text{AM-RE :}\quad&\mathcal{D}_{\mathcal{R}}(\mathbf{z}) = \mathrm(|\mathbf{z}|-\mathcal{R}(|\mathbf{z}|))_+ \odot \mathrm{sign}(\mathbf{z}),
\label{eq:residual}
\end{align}
where $(\cdot)_+ = \max(\cdot,0)$ represents ReLU.

With existing approaches for constructing Lipschitz continuous DNNs, the mappings $\mathcal{S}$ and $\mathcal{R}$ can be readily designed to be Lipschitz continuous.
Nevertheless, the entire systems $\mathcal{D}_{\mathcal{S}}$ and $\mathcal{D}_{\mathcal{R}}$ are generally not Lipschitz continuous. 
To ensure Lipschitz continuity of AMs, a special technique is required, which we introduce next.

\section{Proposed Method}

In this section, we establish a sufficient condition that guarantees the Lipschitz continuity of AMs.
We then propose Lipschitz-continuous variants for AM-SE and AM-RE and derive theoretical bounds on their Lipschitz constants.
To emphasize the necessity of our method, 
we first show that AMs are generally not Lipschitz continuous.

\subsection{AMs are not Lipschitz Continuous}
The AMs $\mathcal{D}_{\mathcal{A}}$, $\mathcal{D}_{\mathcal{S}}$ and $ \mathcal{D}_{\mathcal{R}}$ in Eqs.~\eqref{eq:amplitude}--\eqref{eq:residual} are not Lipschitz continuous in general, even when the amplitude-modifying parts $\mathcal{A}$, $\mathcal{S}$ and $ \mathcal{R}$ are Lipschitz continuous.
We demonstrate this fact through the following two examples:
\begin{example}[Bias]
\label{ex:bias}
The mapping $\mathcal{A}:\mathbb{R}_+\to \mathbb{R}_+:x\mapsto x + 1$ is 1-Lipschitz continuous. However, $\mathcal{D}_{\mathcal{A}}:z\mapsto(|z|+1)\cdot\mathrm{sign}(z)$ is not Lipschitz continuous.%
\footnote{Let
$z\in\mathbb{C}$ such that $|z|=\epsilon>0$ and let $w=-z$, then $|\mathcal{D}_{\mathcal{A}}(z)-\mathcal{D}_{\mathcal{A}}(w)|/|z-w| = 2(\varepsilon+1) / (2\varepsilon) \to +\infty$ with $\varepsilon \to +0$. }
\end{example}
\begin{example}[Permutation]
\label{ex:permutation}
The mapping $\mathcal{A}:\mathbb{R}_+^2\to \mathbb{R}_+^2:(x_1,x_2)\mapsto (x_2,x_1)$ is 1-Lipschitz continuous.
However, $\mathcal{D}_{\mathcal{A}}:\mathbf{z}\mapsto\mathcal{{A}}(|\mathbf{z}|)\odot\mathrm{sign}(\mathbf{z})$ is not Lipschitz continuous.%
\footnote{Let $\mathbf{z}=(\varepsilon,1)\in\mathbb{C}^2$, $\mathbf{w}=(-\varepsilon,1)\in\mathbb{C}^2$, where $\varepsilon >0$. 
Then $\|\mathcal{D}_{\mathcal{A}}(\mathbf{z})-\mathcal{D}_{\mathcal{A}}(\mathbf{w})\|_2/\|\mathbf{z}-\mathbf{w}\|_2 =  2 / (2\varepsilon)\to +\infty$ with $\varepsilon \to +0$.}
\end{example}

These examples lead to the following question: \textit{How can we impose Lipschitz continuity on AMs?}
We provide an answer to this question in the following subsection.

\subsection{LipsAM: Lipschitz-continuous Amplitude Modifier}
Aiming to provide a tool to enforce Lipschitz continuity for AMs, 
we introduce some assumptions on the mapping $\mathcal{A}$ in Eq.~\eqref{eq:amplitude}:
\begin{asm}
\label{asm:assumption}
    For any $\mathbf{x}$, $\mathbf{y}\in\mathbb{R}_+^N$ and for all $n\in \{1,\ \ldots,N\}$, the mapping $\mathcal{A}:\mathbb{R}_+^N\to \mathbb{R}_+^N$ satisfies the following two conditions with some constants $L_1, L_2\geq 0$.
\begin{enumerate}
    \item $\mathcal{A}$ is $L_1$-Lipschitz continuous, i.e.,
    \begin{equation}
        \quad\|\mathcal{A}(\mathbf{x}) - \mathcal{A}(\mathbf{y}) \|_2 \leq L_1 \|\mathbf{x}-\mathbf{y}\|_2.
        \label{eq:cond1}
    \end{equation}
    \item For each element, $\mathcal{A}(\mathbf{x})$ is bounded by $L_2\mathbf{x}$, i.e.,
    \begin{equation}
    0\leq (\mathcal{A}(\mathbf{x}))_n \leq L_2 x_n.
        \label{eq:cond2}
    \end{equation}
\end{enumerate}
\end{asm}
\noindent
That is, in addition to Lipschitz continuity, 
the output of $\mathcal{A}$ is required to be element-wise bounded as in the second condition.

We show that the AM with such $\mathcal{A}$ is Lipschitz bounded.
\begin{thm}
\label{lem:cond}
Let $\mathcal{D}_{\mathcal{A}}:\mathbb{C}^N\to\mathbb{C}^N:\mathbf{z}\mapsto \mathcal{A}(|\mathbf{z}|)\odot\mathrm{sign}(\mathbf{z})$, and $\mathcal{A}:\mathbb{R}_+^N\to\mathbb{R}_+^N$ satisfy Eqs.~\eqref{eq:cond1}  and \eqref{eq:cond2} in Assumption~\ref{asm:assumption}.
Then, $\mathcal{D}_{\mathcal{A}}$ is Lipschitz continuous, and its Lipschitz constant is bounded as
\begin{equation}
    \mathrm{Lip}(\mathcal{D}_{\mathcal{A}})\leq \max(L_1, L_2).
\end{equation}
\begin{proof}
Take arbitrary $\mathbf{z}$, $\mathbf{w} \in \mathbb{C}^N$ and represent them in the polar form as $\mathbf{z} = \mathbf{x}\odot \exp(\mathrm{i}\boldsymbol{\phi})$, $\mathbf{w} = \mathbf{y}\odot \exp(\mathrm{i}\boldsymbol{\psi})$, respectively,
where $\mathbf{x}$, $\mathbf{y} \in \mathbb{R}_+^N$, $\boldsymbol{\phi}$, $\boldsymbol{\psi} \in [0,2\pi)^N$, $\mathrm{i}=\sqrt{-1}$, and $\exp(\cdot)$ is the element-wise exponential.
Using Eqs.~\eqref{eq:cond1} and \eqref{eq:cond2}, we obtain
$\|\mathcal{D}_{\mathcal{A}}(\mathbf{z}) - \mathcal{D}_{\mathcal{A}}(\mathbf{w})\|_2^2 
= \|\mathcal{A}(\mathbf{x}) - \mathcal{A}(\mathbf{y})\|_2^2 
+ 2\sum_{n=1}^N (\mathcal{A}(\mathbf{x}))_n (\mathcal{A}(\mathbf{y}))_n \bigl(1 - \cos(\phi_n - \psi_n)\bigr) 
\leq (\max(L_1,L_2))^2 \|\mathbf{z} - \mathbf{w}\|_2^2$.
\end{proof}
\end{thm}
This result shows that an AM $\mathcal{D}_\mathcal{A}$ is Lipschitz continuous when the mapping $\mathcal{A}$ acting on the amplitude satisfies Assumption~\ref{asm:assumption}.
We refer to such a Lipschitz-continuous AM as \textit{LipsAM}.

\subsection{Proposed Architectures: LipsAM-SE and LipsAM-RE}
Using the above result, we propose 
LipsAM-SE and LipsAM-RE by introducing some layers into the architectures in Eqs.~\eqref{eq:signal} and \eqref{eq:residual}, so that the DNNs satisfy the conditions in Assumption~\ref{asm:assumption}:
\begin{align}
\!\!\text{LipsAM-SE : }\;&\mathcal{D}_{\mathcal{S}}^\mathrm{(Lips)}
(\mathbf{z}) = (\min(\mathcal{S}(|\mathbf{z}|), |\mathbf{z}|))_+ \odot \mathrm{sign}(\mathbf{z}),
\label{eq:signallim}\\
\!\!\text{LipsAM-RE : }\;&
\mathcal{D}_{\mathcal{R}}^\mathrm{(Lips)}(\mathbf{z}) = \mathrm(|\mathbf{z}|-(\mathcal{R}(|\mathbf{z}|))_+)_+ \odot \mathrm{sign}(\mathbf{z}).
\label{eq:residuallim}
\end{align}
These architectures are graphically shown in Fig.~\ref{fig:architecture}.

Since the blue layers in Fig.~\ref{fig:architecture} (the element-wise $\min$ in Eq.~\eqref{eq:signallim} and $(\cdot)_+$ applied just after $\mathcal{R}$ in Eq.~\eqref{eq:residuallim}) ensure that the outputs are smaller than the inputs, Eq.~\eqref{eq:cond2} is satisfied with $L_2 = 1$.
In addition, by evaluating the Lipschitz constants of their amplitude-modifying parts, $L_1$ in Eq.~\eqref{eq:cond1}, we establish their Lipschitz bounds as follows.

\begin{thm}
\label{thm:signallim}
Assume $\mathcal{S}:\mathbb{R}_+^N \to \mathbb{R}^N$ in Eq.~\eqref{eq:signallim} and $\mathcal{R}:\mathbb{R}_+^N \to \mathbb{R}^N$ in Eq.~\eqref{eq:residuallim} are Lipschitz continuous.
Then, $\mathcal{D}_{\mathcal{S}}^\mathrm{{(Lips)}}:\mathbb{C}^N \to \mathbb{C}^N$ in Eq.~\eqref{eq:signallim} and $\mathcal{D}_{\mathcal{R}}^\mathrm{{(Lips)}}:\mathbb{C}^N \to \mathbb{C}^N$ in Eq.~\eqref{eq:residuallim} are Lipschitz continuous, and their Lipschitz constants are bounded as
\begin{align}
    \mathrm{Lip}(\mathcal{D}_{\mathcal{S}}^\mathrm{{(Lips)}})&\leq \sqrt{(\mathrm{Lip}(\mathcal{S}))^2+1},\\
    \mathrm{Lip}(\mathcal{D}_{\mathcal{R}}^\mathrm{{(Lips)}})&\leq \mathrm{Lip}(\mathcal{R})+1.
\end{align}
\begin{proof}
We show the bound on the Lipschitz constant of the amplitude-modifying parts.
The operator $(\min(\mathcal{S}(\cdot),\cdot))_+$ is decomposed as
\begin{align*}
\Bigl(\mathbb{R}_+^{2N}\ni \boldsymbol{\xi}\mapsto ((\min(\xi_n, \xi_{N+n}))_+)_{n=1}^N\in \mathbb{R}^N_+
\Bigr)\circ \begin{pmatrix}
    \mathcal{S}\\\mathrm{Id}
    \end{pmatrix},
\end{align*}
where $\mathrm{Id}$ is the identity operator.
The former part is $1$-Lipschitz continuous, and 
the latter part is $\sqrt{(\mathrm{Lip}(\mathcal{S}))^2+1}$-Lipschitz continuous, thus $1\leq \mathrm{Lip}((\min(\mathcal{S}(\cdot),\cdot))_+)\leq 1\times\sqrt{(\mathrm{Lip}(\mathcal{S}))^2+1}$.
On the other hand, we have $\|(\mathbf{x}-(\mathcal{R}(\mathbf{x}))_+) - (\mathbf{y}-(\mathcal{R}(\mathbf{y}))_+)\|_2 \leq \|\mathbf{x}-\mathbf{y}\|_2+\|(\mathcal{R}(\mathbf{x}))_+-(\mathcal{R}(\mathbf{y}))_+\|_2$ for any $\mathbf{x}$,  $\mathbf{y}\in\mathbb{R}^N$ and also $(\cdot)_+$ is 1-Lipschitz continuous.
Therefore, $1\leq \mathrm{Lip}((\cdot-(\mathcal{R}(\cdot))_+)_+)\leq (1+\mathrm{Lip}(\mathcal{R}))\times 1$.
These facts with Theorem \ref{lem:cond} prove Theorem \ref{thm:signallim}.
\end{proof}
\end{thm}
The proposed approach is straightforward to implement: Once Lipschitz continuous DNNs, $\mathcal{S}$ and $\mathcal{R}$, are trained, 
one only needs to insert the blue layers in Fig.~\ref{fig:architecture} to enforce the AMs to be LipsAMs.

\begin{figure}
    \centering
    \includegraphics[scale=0.59]{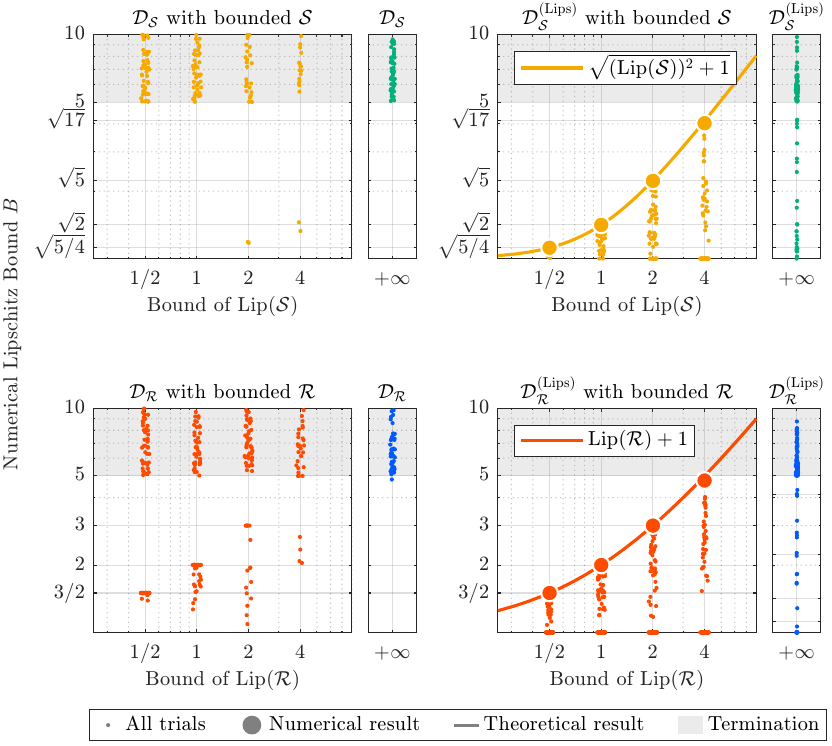}
    \caption{Numerical estimate of the value $B$ in Eq.~\eqref{eq:B} for each architecture. 
    Upper rows are that of AM-SE in Eq.~\eqref{eq:signal} and LipsAM-SE in Eq.~\eqref{eq:signallim}.
    Lower rows are that of AM-RE in Eq.~\eqref{eq:residual} and LipsAM-RE in Eq.~\eqref{eq:residuallim}.
    Dots indicate results of 100 trials, and maximum result is marked with large circle. 
    Solid lines show the theoretical bound in Theorem \ref{thm:signallim}. 
    Areas exceeding termination threshold are darkened.}
    \label{fig:bounds}
\end{figure}

\subsection{Numerical Validation of Lipschitz Bound}
Here, we numerically validate Theorem \ref{thm:signallim} by approximately computing the following quantity for each architecture:
\begin{align}
    B = \sup_{\mathbf{z}\in\mathbb{C}^N,\;\boldsymbol{\theta}\in\boldsymbol{\Theta}} \left\|\mathbf{J}_{\mathcal{D}}(\mathbf{z};{\boldsymbol{\theta}})\right\|_{\rm op},
    \label{eq:B}
\end{align}
where
$\mathcal{D} \in \{\mathcal{D}_\mathcal{S},\mathcal{D}_\mathcal{R},\mathcal{D}_\mathcal{S}^\mathrm{(Lips)},\mathcal{D}_\mathcal{R}^\mathrm{(Lips)}\}$,
and $\boldsymbol{\theta}$ represents the trainable parameters of $\mathcal{S}$ and $\mathcal{R}$. 
$\mathbf{J}_f(\mathbf{x}) = \left[\partial f_i/\partial x_j(\mathbf{x})\right]_{ij} \in \mathbb{R}^{K \times K}$ is the Jacobian matrix%
\footnote{A complex-valued function $f:\mathbb{C}^K\to\mathbb{C}^K$ can be identified with a real-valued function $\tilde f:\mathbb{R}^{2K}\to\mathbb{R}^{2K}$, and then its Jacobian matrix can be defined as the same manner as for the real-valued mappings.} of $f:\mathbb{R}^K\to \mathbb{R}^K$ at $\mathbf{x}\in\mathbb{R}^K$, and $\|\cdot \|_\mathrm{op}$ is the operator norm (i.e., the maximum singular value).
In general, if $f$ is an $L$-Lipschitz continuous smooth function, then $\|\mathbf{J}_f(\mathbf{x})\|_{\rm op}\leq L$ holds for all $\mathbf{x}\in\mathbb{R}^K$ \cite{virmaux_lipschitz_2018}. 
Therefore, if a DNN architecture is $L$-Lipschitz continuous by design, then $B \leq L$ must hold.

For $\mathcal{S}$ and $\mathcal{R}$, 1-Lipschitz CNNs were constructed using orthogonal convolution layers \cite{li_preventing_2019}, and the outputs were scaled by 0.5, 1, 2, and 4.
Ordinary CNNs without constraints, whose Lipschitz constants are not bounded (i.e., $+\infty$), were also tested for comparison.
$\mathrm{SoftPlus}(\cdot)=\log(1+\exp(\cdot))$ was used for the activation function.\footnote{
We used the smooth activation function because $B$ in Eq.~\eqref{eq:B} involves the Jacobian matrix of a DNN, 
and its gradient for optimization contains the second-order derivatives of the activation function.
Note that such smoothness of activation functions is a requirement only for this experiment. %
}
The input vector $\mathbf{z} \in \mathbb{C}^N$ was $4\times4$ complex-valued one-channel image, i.e., $N=16$. 
The kernel size for the convolution layer was $3\times 3$. 
The number of channels in each convolution layer was 3, except for that of the final one, which was 1.
Adam was used for optimization, 
with a learning rate of 0.1 and for a maximum of 1000 iterations.
The optimization was terminated once $B$ exceeded 5 to save computational cost.
The operator norm was computed via the power iteration to utilize automatic differentiation \cite{miyato_spectral_2018}.
The initial values for $\mathbf{z}$ and $\boldsymbol{\theta}$ were randomly chosen with 100 seeds.

As shown in Fig.~\ref{fig:bounds}, the value of $B$ for the conventional AMs ($\mathcal{D}_\mathcal{S}$ and $\mathcal{D}_\mathcal{R}$) exceeded the termination threshold (set to 5).
In contrast, the values $B$ for LipsAMs ($\mathcal{D}_\mathcal{S}^\mathrm{(Lips)}$ and $\mathcal{D}_\mathcal{R}^\mathrm{(Lips)}$) with Lipschitz-bounded DNNs were tightly bounded by the theoretical bound established in Theorem~\ref{thm:signallim}, which confirmed our results.

\section{Application to Plug-and-Play \\ Speech Dereverberation}

We applied the proposed LipsAMs to PnP speech dereverberation as an example of their audio signal processing application.

\subsection{Problem Setting and Plug-and-Play Algorithm}
\label{subsubsec:problem}
Let a signal be observed through a noisy reverberant system $\mathbf{y}=\mathbf{H}\mathbf{s} + \mathbf{n}\in\mathbb{R}^T$, where $T$ is the signal length, $\mathbf{s}\in \mathbb{R}^T$ is the clean signal aimed to be recovered, $\mathbf{H}\in\mathbb{R}^{T\times T}$ is a known convolution matrix for the impulse response $\mathbf{h}\in\mathbb{R}^T$, and $\mathbf{n}\in\mathbb{R}^T$ is the additive Gaussian noise.
The signal recovery problem can be formulated as %
\begin{equation}
\min_{\mathbf{x}\in\mathbb{R}^T} \left(\lambda\Omega(\mathbf{G}\mathbf{x})+(1/2)\|\mathbf{Hx}-\mathbf{y}\|_2^2\right),
\label{eq:generalFormulation}
\vspace{-4pt}
\end{equation}
where $\mathbf{x}\in\mathbb{R}^T$ is the recovered signal, $\mathbf  G\in\mathbb{C}^{N\times T}$ denotes the STFT operator, and $\lambda>0$ is the weight of the regularization term.
The regularization function $\Omega$ incorporates prior knowledge on the spectrogram, such as $\ell_1$-norm for inducing sparsity.
The proximal algorithms \cite{boyd_distributed_2010} can handle this problem via the proximity operator, $\mathrm{prox}_{\mu \Omega}(\mathbf{v}) = \arg\min_\mathbf{u} \left(\Omega(\mathbf{u}) + (1/2\mu) \|\mathbf{u} - \mathbf{v}\|_2^2 \right)$, where $\mu>0$ is a parameter.
In PnP algorithms, a DNN-based denoiser for audio signals is used in place of the proximity operator, which implicitly incorporates data-driven priors into iterative optimization algorithms \cite{matsumoto2024determined,tanaka_applade_2022}.
Note that the convergence of such algorithms typically requires a non-expansive DNN, i.e., its Lipschitz constants should be 1 or less \cite{ryu_plug-and-play_2019,pesquet_learning_2021}.

In this paper, we use the following algorithm derived from the alternating direction method of multipliers (ADMM) \cite{boyd_distributed_2010},
\begin{align}
\hspace{-15pt}\left\lfloor\;
\begin{aligned}
\mathbf{x} &\gets (\mathbf{H}^{\mathsf{H}} \mathbf{H} + \mathbf{G}^{\mathsf{H}} \mathbf{G})^{-1}
\bigl( \mathbf{H}^{\mathsf{H}} (\mathbf{u} - \boldsymbol{\xi}_1) + \mathbf{G}^\mathsf{H} (\mathbf{v} - \boldsymbol{\xi}_2) \bigr), \hspace{-30pt}\\
\mathbf{u} &\gets \mathrm{prox}_{(1/2\lambda)\|\cdot\|_2^2}( \mathbf{H} \mathbf{x} + \boldsymbol{\xi}_1 -\mathbf{y})+\mathbf{y}, \\
\mathbf{v} &\gets  \mathcal{D}\bigl( \mathbf{B} \mathbf{x} + \boldsymbol{\xi}_2 \bigr), \\
(\boldsymbol{\xi}_1, \boldsymbol{\xi}_2) &\gets (\boldsymbol{\xi}_1, \boldsymbol{\xi}_2) + (\mathbf{A} \mathbf{x} - \mathbf{u},\mathbf{B} \mathbf{x} - \mathbf{v}),
\end{aligned}
\right.
\label{eq:admm}
\end{align}
where $\mathbf{u}$, $ \boldsymbol{\xi}_1\in\mathbb{C}^{T}$, and $\mathbf{v}$, $ \boldsymbol{\xi}_2\in\mathbb{C}^{N}$.
Note that, by using a tight window for STFT (i.e., $\mathbf{G}^\mathsf{H}\mathbf{G}=\mathbf{I}$, where $\mathbf{I}$ is the identity matrix \cite{janssen_characterization_2002}) and assuming the circular convolution, the matrix inversion in the 1st line can be efficiently computed using the fast Fourier transform as $\mathsf{iFFT}(\mathsf{FFT}(\cdot)\oslash( |\mathsf{FFT}(\mathbf{h})|^2+\mathbf{1}))$, where $\mathbf{1}\in\{1\}^T$.

\begin{figure}
    \centering
    \includegraphics[scale=0.59]{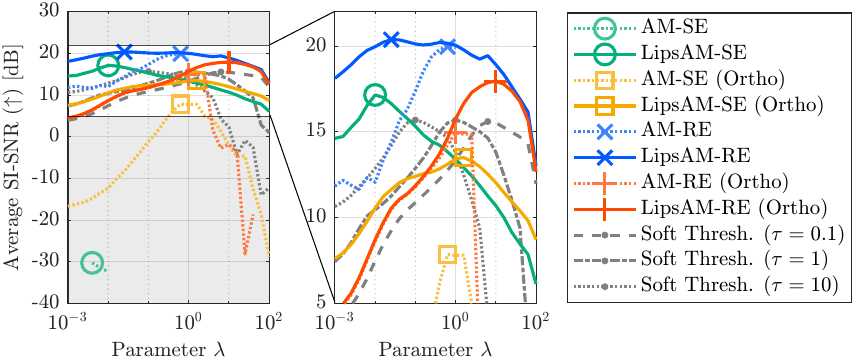}
    \caption{Parameter $\lambda$ vs SI-SNR after 500 iterations.
    Solid colored lines represents the proposed LipsAMs,
    and dotted colored lines represent the conventional AMs.
    Use of orthogonal layers is indicated by (Ortho).
    Gray lines show $\ell_1$-norm-based method.
    Best parameters for each model are shown by the markers.
    A close-up view of left is on the right.
    Missing points indicate that algorithms diverged.}
    \label{fig:parameters}
\end{figure}

\subsection{Experimental Settings}
\textbf{Architecture: }
The two DNNs based on AM-SE in Eq.~\eqref{eq:signal} and AM-RE in Eq.~\eqref{eq:residual} were trained, and the proposed LipsAM-SE in Eq.~\eqref{eq:signallim} and LipsAM-RE in Eq.~\eqref{eq:residuallim} are constructed by inserting the layers we introduced to each of them.
The architecture for their learnable part $\mathcal{S}$ and $\mathcal{R}$ includes one-dimensional convolution (Conv1D) layers, where frequency bins are treated as channels. 
It consists of three Conv1D layers with a kernel size of 5, and we used a leaky ReLU activation with a slope of 0.1. 
The intermediate feature dimension was set to 512 channels.
In addition to the standard Conv1D, we also tested orthogonal Conv1D layers \cite{trockman_orthogonalizing_2020}, which restrict $\mathcal{S}$ and $\mathcal{R}$ to be 1-Lipschitz continuous.

\textbf{Noisy Signal for Training: }
\label{sec:noise}
The denoisers were trained on the Gaussian denoising task on speech signals.
During training, Gaussian noise was added to the clean signals in a signal-to-noise ratio (SNR) uniformly sampled between 20 and 40\,dB, and the DNNs were optimized to suppress this noise. 
The audio signals were downsampled to 8\,kHz.
The time-frequency representation was obtained using the STFT with the Parseval-tight window made from the Hann window.
The window length was 512 samples, and the hop size was 256 samples. 
The signals were truncated to 32 time frames in the time-frequency domain.
The loss function was the negative SNR in the time domain. 
For training data, the \texttt{train-clean-100} subset of LibriTTS-R \cite{koizumi_libritts-r_2023} was used, where 10\% of that was used for validation.
Each DNN was trained for a maximum of 20 epochs and the best model on the validation set was used. 
Adam with a learning rate of $1.0 \times 10^{-4}$ was used and the batch size was set to 32.

\textbf{Reverberant Signals for Recovery: }
\label{sec:rev}
The source signal $\mathbf{s}$ and the impulse response $\mathbf{h}$ in Eq.~\eqref{eq:generalFormulation} were randomly chosen from the \texttt{test-clean} subset of LibriTTS\_R \cite{koizumi_libritts-r_2023} and the BUT reverb database \cite{szoke_building_2019}, respectively.
The noise level of $\mathbf{n}$ was set at 30 dB to $\mathbf{H}\mathbf{s}$.  The other settings were the same as the data used for training.

\subsection{Results}

\textbf{Parameter Selection: }
We searched for the appropriate value of $\lambda$, i.e., the weight of the regularization term in Eq.~\eqref{eq:generalFormulation}.
Five noisy and reverberant signals were generated as described in Section \ref{sec:rev}, and the PnP algorithm was run for 500 iterations for each DNN.
The value of $\lambda$ ranged from $10^{-3}$ to $10^{2}$ with 26 logarithmically spaced values. The $\ell_1$-norm-based method was compared as a baseline, in which the soft-thresholding operator $\mathrm{SoftThresh}_\tau =(|\cdot|-\tau)_+\odot \mathrm{sign}(\cdot)$ is used in place of $\mathcal{D}$ in Eq.~\eqref{eq:admm}.

Fig.~\ref{fig:parameters} shows the average scale-invariant SNR (SI-SNR) for each method after 500 iterations.
The conventional AMs were not robust to the choice of $\lambda$.
Specifically, AM-SE tended to diverge and caused NaN for most $\lambda$.
In contrast, LipsAMs (solid lines) successfully avoided such divergence and ran stably.
In particular, LipsAM-RE yielded the best performance for all settings of $\lambda$.

\begin{table}
\footnotesize
\centering
\tabcolsep = 3pt
\caption{Evaluation results with best parameter $\lambda$ marked in Fig.~\ref{fig:parameters}. N/A indicates that algorithm diverged. Bold and underlined numbers indicate the best and second-best values, respectively.}
\vspace{4pt}
\label{tab:results}
\begin{tabular}{lll|rrrr} \hline
\multicolumn{3}{l|}{Denoiser $\mathcal{D}$ in Eq.~\eqref{eq:admm}} & SI-SNR ($\uparrow$) & PESQ ($\uparrow$) & STOI ($\uparrow$) & ViSQOL ($\uparrow$) \\ \Xhline{1pt}
\rowcolor{gray!10}    AM   & \!\!\!-SE   &                           & N/A & N/A & N/A & N/A               \\
\rowcolor{gray!10} LipsAM  & \!\!\!-SE   &           & $16.61$ & $2.91$ & $0.91$ & $3.64$             \\
\rowcolor{white}      AM   & \!\!\!-SE   & (Ortho)                      & $9.54$ & $2.30$ & $0.88$ & $3.10$            \\
\rowcolor{white}   LipsAM  & \!\!\!-SE   & (Ortho)      & $14.44$ & $2.68$ & $0.93$ & $3.75$             \\ 
\rowcolor{gray!10}   AM    & \!\!\!-RE &                             & $17.98$ & $\mathbf{3.21}$ & $\underline{0.97}$ & $\textbf{4.21}$                                        \\
\rowcolor{gray!10} LipsAM  & \!\!\!-RE &             & $\mathbf{20.57}$ & $\underline{3.14}$ & $\mathbf{0.97}$ & $\underline{4.21}$                                        \\
\rowcolor{white}       AM  & \!\!\!-RE &    (Ortho)                        & N/A & N/A & N/A & N/A                      \\
\rowcolor{white}   LipsAM  & \!\!\!-RE &    (Ortho)        & $\underline{18.64}$ & $2.90$ & $0.95$ & $3.94$             \\ \hline
\rowcolor{gray!10} \multicolumn{3}{l|}{Soft Thresh. $(\tau=0.1)$}      & $17.34$ & $2.95$ & $0.96$ & $3.89$             \\
\rowcolor{gray!10} \multicolumn{3}{l|}{Soft Thresh. $(\tau=1)$}        & $16.70$ & $2.91$ & $0.95$ & $3.75$             \\
\rowcolor{gray!10} \multicolumn{3}{l|}{Soft Thresh. $(\tau=10)$}       & $16.70$ & $2.91$ & $0.95$ & $3.75$             \\ \hline
\end{tabular}
\vspace{2pt}
\end{table}

\begin{figure}
    \centering
    \includegraphics[scale=0.62]{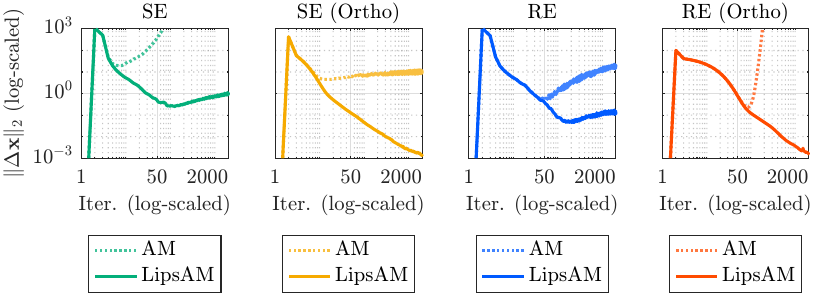}
    \vspace{-2pt}
    \caption{Transition of amount of update $\|\Delta \mathbf{x}\|_2$.
    In each box, results of LipsAMs with best $\lambda$ in Fig.~\ref{fig:parameters} were compared with that of AMs with the same $\lambda$ as LipsAMs.
    }
    \vspace{-2pt}
    \label{fig:convergence}
\end{figure}

\textbf{Evaluation: } We further evaluated the methods using 10 additional signals with additional metrics (PESQ\cite{torcoli_objective_2021},
STOI\cite{taal_short-time_2010} and ViSQOL\cite{hines_visqol_2015}). 
The number of iterations was increased to 2000 to see stability. 
Table \ref{tab:results} shows that LipsAMs performed with higher stability compared to AMs. 
Specifically, AM-SE and AM-RE (Ortho) diverged and resulted in NaN, and LipsAM-SE and LipsAM-RE (Ortho) avoided such failures in this experiment.

Fig.~\ref{fig:convergence} shows the amount of update $\|\Delta \mathbf{x}\|_2^{[k]} = \|\mathbf{x}^{[k]} - \mathbf{x}^{[k-1]}\|_2$
that measures how much the variable $\mathbf{x}$ changed at each
iteration $k=1,\ldots, 2000$, which is supposed to converge to $0$ if the algorithm converges.
LipsAMs clearly contributed to the decrease of $\|\Delta \mathbf{x}\|_2^{[k]}$ over iterations. The orthogonal layers, which serve DNNs with bounded Lipschitz constants, further improved the convergence.

\vspace{-2pt}
\section{Conclusion}
This paper proposed LipsAM, the Lipschitz-continuous DNNs that modify the amplitude of the input.
The methodology we derived, together with the bound on the Lipschitz constant, can provide useful tools for the theoretical analysis of these DNNs. 
Furthermore, we demonstrated the effectiveness of our technique in a PnP-based speech signal recovery task. 
Future works include extending the proposed approach to time-frequency-masking-based DNNs and establishing convergence guarantees for the PnP methods.

\bibliographystyle{IEEEtran}
\bibliography{reference_clean}

\end{document}